\documentclass{article}
\usepackage{setspace}
\usepackage{psfig}
\usepackage[super]{natbib}

\begin{document}

\begin{center}
{\Large \bf 
\noindent
{The suppression of star formation by powerful active galactic nuclei}}
\end{center}
\vspace{4mm}

\singlespacing
\begin{center}
\noindent
{\parbox{\textwidth}{\raggedright M.J.~Page,$^{1}$
M.~Symeonidis,$^{1}$
J.D.~Vieira,$^{2}$
B.~Altieri,$^{3}$
A.~Amblard,$^{4}$
V.~Arumugam,$^{5}$
H.~Aussel,$^{6}$
T.~Babbedge,$^{7}$
A.~Blain,$^{8}$
J.~Bock,$^{2,9}$
A.~Boselli,$^{10}$
V.~Buat,$^{10}$
N.~Castro-Rodr\'iguez,$^{11,12}$
A.~Cava,$^{13}$
P.~Chanial,$^{6}$
D.L.~Clements,$^{7}$
A.~Conley,$^{14}$
L.~Conversi,$^{3}$
A.~Cooray,$^{15,2}$
C.D.~Dowell,$^{2,9}$
E.N.~Dubois,$^{16}$
J.S.~Dunlop,$^{5}$
E.~Dwek,$^{17}$
S.~Dye,$^{18}$
S.~Eales,$^{19}$
D.~Elbaz,$^{6}$
D.~Farrah,$^{16}$
M.~Fox,$^{7}$
A.~Franceschini,$^{20}$
W.~Gear,$^{19}$
J.~Glenn,$^{21,14}$
M.~Griffin,$^{19}$
M.~Halpern,$^{22}$
E.~Hatziminaoglou,$^{23}$
E.~Ibar,$^{24}$
K.~Isaak,$^{25}$
R.J.~Ivison,$^{24,5}$
G.~Lagache,$^{26}$
L.~Levenson,$^{2,9}$
N.~Lu,$^{2,27}$
S.~Madden,$^{6}$
B.~Maffei,$^{28}$
G.~Mainetti,$^{20}$
L.~Marchetti,$^{20}$
H.T.~Nguyen,$^{9,2}$
B.~O'Halloran,$^{7}$
S.J.~Oliver,$^{16}$
A.~Omont,$^{29}$
P.~Panuzzo,$^{6}$
A.~Papageorgiou,$^{19}$
C.P.~Pearson,$^{30,31}$
I.~P\'erez-Fournon,$^{11,12}$
M.~Pohlen,$^{19}$
J.I.~Rawlings,$^{1}$
D.~Rigopoulou,$^{30,32}$
L.~Riguccini,$^{6}$
D.~Rizzo,$^{7}$
G.~Rodighiero,$^{20}$
I.G.~Roseboom,$^{16,5}$
M.~Rowan-Robinson,$^{7}$
M.~S\'anchez Portal,$^{3}$
B.~Schulz,$^{2,27}$
Douglas~Scott,$^{22}$
N.~Seymour,$^{33,1}$
D.L.~Shupe,$^{2,27}$
A.J.~Smith,$^{16}$
J.A.~Stevens,$^{34}$
M.~Trichas,$^{35}$
K.E.~Tugwell,$^{1}$
M.~Vaccari,$^{20}$
I.~Valtchanov,$^{3}$
M.~Viero,$^{2}$
L.~Vigroux,$^{29}$
L.~Wang,$^{16}$
R.~Ward,$^{16}$
G.~Wright,$^{24}$
C.K.~Xu$^{2,27}$ and
M.~Zemcov$^{2,9}$}\vspace{0.4cm}\\
\parbox{\textwidth}{\raggedright $^{1}$Mullard Space Science Laboratory, University College London, Holmbury St. Mary, Dorking, Surrey RH5 6NT, UK\\
$^{2}$California Institute of Technology, 1200 E. California Blvd., Pasadena, CA 91125, USA\\
$^{3}$Herschel Science Centre, European Space Astronomy Centre, Villanueva de la Ca\~nada, 28691 Madrid, Spain\\
$^{4}$NASA, Ames Research Center, Moffett Field, CA 94035, USA\\
$^{5}$Institute for Astronomy, University of Edinburgh, Royal Observatory, Blackford Hill, Edinburgh EH9 3HJ, UK\\
$^{6}$Laboratoire AIM-Paris-Saclay, CEA/DSM/Irfu - CNRS - Universit'e Paris Diderot, CE-Saclay, pt courrier 131, F-91191 Gif-sur-Yvette, France\\
$^{7}$Astrophysics Group, Imperial College London, Blackett Laboratory, Prince Consort Road, London SW7 2AZ, UK\\
$^{8}$Department of Physics \& Astronomy, University of Leicester, University Road, Leicester LE1 7RH, UK\\
$^{9}$Jet Propulsion Laboratory, 4800 Oak Grove Drive, Pasadena, CA 91109, USA\\
$^{10}$Laboratoire d'Astrophysique de Marseille, OAMP, Universit'e Aix-marseille, CNRS, 38 rue Fr'ed'eric Joliot-Curie, 13388 Marseille cedex 13, France\\
$^{11}$Instituto de Astrof{'\i}sica de Canarias (IAC), E-38200 La Laguna, Tenerife, Spain\\
$^{12}$Departamento de Astrof{'\i}sica, Universidad de La Laguna (ULL), E-38205 La Laguna, Tenerife, Spain\\
$^{13}$Departamento de Astrof'isica, Facultad de CC. F'isicas, Universidad Complutense de Madrid, E-28040 Madrid, Spain\\
$^{14}$Center for Astrophysics and Space Astronomy 389-UCB, University of Colorado, Boulder, CO 80309, USA\\
$^{15}$Dept. of Physics \& Astronomy, University of California, Irvine, CA 92697, USA\\
$^{16}$Astronomy Centre, Dept. of Physics \& Astronomy, University of Sussex, Brighton BN1 9QH, UK\\
$^{17}$Observational Cosmology Lab, Code 665, NASA Goddard Space Flight  Center, Greenbelt, MD 20771, USA\\
$^{18}$School of Physics and Astronomy, University of Nottingham, NG7 2RD, UK\\
$^{19}$School of Physics and Astronomy, Cardiff University, Queens Buildings, The Parade, Cardiff CF24 3AA, UK\\
$^{20}$Dipartimento di Astronomia, Universit\`{a} di Padova, vicolo Osservatorio, 3, 35122 Padova, Italy\\
$^{21}$Dept. of Astrophysical and Planetary Sciences, CASA 389-UCB, University of Colorado, Boulder, CO 80309, USA\\
$^{22}$Department of Physics \& Astronomy, University of British Columbia, 6224 Agricultural Road, Vancouver, BC V6T~1Z1, Canada\\
$^{23}$ESO, Karl-Schwarzschild-Str. 2, 85748 Garching bei M"unchen, Germany\\
$^{24}$UK Astronomy Technology Centre, Royal Observatory, Blackford Hill, Edinburgh EH9 3HJ, UK\\
$^{25}$European Space Research and Technology Centre (ESTEC), Keplerlaan 1, 2201 AZ, Noordwijk, The Netherlands\\
$^{26}$Institut d'Astrophysique Spatiale (IAS), b\^atiment 121, Universit'e Paris-Sud 11 and CNRS (UMR 8617), 91405 Orsay, France\\
$^{27}$Infrared Processing and Analysis Center, MS 100-22, California Institute of Technology, JPL, Pasadena, CA 91125, USA\\
$^{28}$School of Physics and Astronomy, The University of Manchester, Alan Turing Building, Oxford Road, Manchester M13 9PL, UK\\
$^{29}$Institut d'Astrophysique de Paris, UMR 7095, CNRS, UPMC Univ. Paris 06, 98bis boulevard Arago, F-75014 Paris, France\\
$^{30}$RAL Space, Rutherford Appleton Laboratory, Chilton, Didcot, Oxfordshire OX11 0QX, UK\\
$^{31}$Institute for Space Imaging Science, University of Lethbridge, Lethbridge, Alberta, T1K 3M4, Canada\\
$^{32}$Department of Astrophysics, Denys Wilkinson Building, University of Oxford, Keble Road, Oxford OX1 3RH, UK\\
$^{33}$CSIRO Astronomy \& Space Science, PO Box 76, Epping, NSW 1710, Australia\\
$^{34}$Centre for Astrophysics Research, University of Hertfordshire, College Lane, Hatfield, Hertfordshire AL10 9AB, UK\\
$^{35}$Harvard-Smithsonian Center for Astrophysics, 60 Garden Street, Cambridge, MA 02138, USA}}
\end{center}
\newpage

\doublespacing

{\bf
\noindent
The old, red stars which constitute the bulges of galaxies, and the
massive black holes at their centres, are the relics of a period in
cosmic history when galaxies formed stars at remarkable rates and 
active galactic
nuclei (AGN) shone brightly from accretion onto black holes.
It is
widely suspected, but unproven, 
that the tight correlation in mass of the black hole
and stellar components\cite{haring04} results from the AGN quenching
the surrounding star formation as it approaches its peak 
luminosity\cite{silk98, fabian99, king10}. X-rays trace emission from AGN 
unambiguously\cite{brandt05},
while powerful star-forming galaxies are usually dust-obscured and are brightest at infrared-submillimetre
wavelengths\cite{sanders96}. 
Here we report observations in the submillimetre and
X-ray which show that rapid star formation was common in the host
galaxies of AGN when the Universe was 2--6 Gyrs old, but that the most
vigorous star formation is not observed around black holes above an
X-ray luminosity of $10^{44}$ erg~s$^{-1}$.  This suppression of star
formation in the host galaxies of powerful AGN is a key prediction of
models in which the AGN drives a powerful
outflow\cite{dimatteo05,springel05,sijacki07}, expelling the
interstellar medium of its host galaxy and transforming the galaxy's
properties in a brief period of cosmic time.}

\vspace{2mm}

Measuring star formation in galaxies containing powerful active
galactic nuclei has long been a problem, because the radiation from
the AGN outshines that from star formation in almost all wavebands. Of
all parts of the electromagnetic spectrum, the far-infrared to
millimetre waveband offers the best opportunity to measure star
formation in galaxies hosting AGN because, in contrast to strongly star-forming
galaxies, AGN emit comparatively little radiation at these 
wavelengths\cite{hatziminaoglou10}.  The combination of deep X-ray and
submillimetre observations therefore offers the best prospects for
studying the association of star formation and accretion during the
$1<z<3$ epoch when star formation and black hole growth in massive
galaxies were at their most vigorous. 

The X-ray catalogue of the {\em Chandra} Deep Field North
(hereafter CDF-N) derives from a series of observations
made with the {\em Chandra} X-ray observatory with a total of 2Ms
exposure time\cite{alexander03}. We restrict the sample to those
sources detected in the most penetrating 2--8~keV band to minimize the
influence of obscuration on our results, and we further limit the
sample to those sources (64\%) for which spectroscopic redshifts are
available in the literature\cite{trouille08,
  barger08}.
2--8~keV
luminosities were calculated assuming that AGN X-ray spectra are power
laws of the form\cite{mateos05} $S_{\nu} \propto \nu^{-0.9}$; the 
luminosities are not corrected for absorption intrinsic to the AGN or 
their host galaxies. In
order to restrict the X-ray sample to AGN we have discarded any
sources with 2--8~keV luminosity
$L_{X} < 10^{42}$~erg~s$^{-1}$. 
SPIRE\cite{griffin10} observations of the CDF-N were carried out in
October 2009 as part of the HerMES programme\cite{oliver10}. 
Maps and source catalogues at 250, 350 and 500~$\mu$m were
constructed\cite{smith10}. At the depth of the SPIRE
maps, the dominant source of uncertainty in the maps is confusion
noise due to the high sky density of sources.  For cross-matching
with the {\em Chandra} source catalogue\cite{alexander03} we chose
the 250~$\mu$m catalogue, which has the most precise positions, and we
used only sources with 250~$\mu$m flux densities $>$ 18~mJy, which corresponds
to a signal/noise ratio $>$3 when the effects of confusion are
included\cite{smith10}. X-ray sources were matched to 250~$\mu$m
sources within 6 arcseconds, corresponding to approximately 95\%
confidence in the 250~$\mu$m positions. The detection statistics are given 
in Table \ref{tab:detections}. The expected level of spurious
associations between X-ray and 250~$\mu$m sources was calculated from
the sky density of 250~$\mu$m sources in 10 to 30 arcsecond-radius
annular regions around the X-ray source positions, and is
reported in Table 1.

\begin{table}
\begin{center}
\begin{tabular}{lcccc}
Region of (z,L) parameter space&Number&Number of AGN&Expected number of&Fraction of AGN \\
&of AGN&associated with&spurious associations&associated with\\
&&250~$\mu$m sources&&250~$\mu$m sources\\
\hline\\
all $z$,\ \ \ \ \ \ \ \ 
$10^{42} < L_{X} < 10^{45}$~erg~s$^{-1}$&176&24&2.1&14$\pm$3 ($_{-5}^{+6}$) \%\\
$1<z<3$, $10^{43} < L_{X} < 10^{44}$~erg~s$^{-1}$&44&11&0.5&25$^{+8}_{-7}$ ($_{-12}^{+15}$) \%\\
$1<z<3$, $10^{44} < L_{X} < 10^{45}$~erg~s$^{-1}$&21&0&0.2&$<5$ ($<13$) \%\\
&&&&\\
\end{tabular}
\end{center}
\caption{250~$\mu$m detection statistics in the various parts of parameter
  space. The first row corresponds to the entire sample of
  secure AGN in the CDF-N, while the second and third rows correspond to the
  regions enclosed within the blue dashed lines in Fig. \ref{fig:z_Lx}. 
Confidence intervals on the fraction of AGN associated with 250~$\mu$m 
sources are given at 68\%, with 95\% intervals enclosed in brackets. 
It should be
  noted that there is one case of two AGN being associated with the
  same 250~$\mu$m source. The two AGN have very similar spectroscopic
  redshifts, and both have X-ray luminosities between $10^{43}$ and
  $10^{44}$~erg~s$^{-1}$. Although the two AGN cannot be resolved at 250~$\mu$m,
  source extraction using X-ray and 24~$\mu$m positions as 
  priors\cite{roseboom10} indicates that both AGN are 5$\sigma$ sources at
  250~$\mu$m. }
\label{tab:detections}
\vspace{0.5cm}
\end{table}

\begin{figure}
\begin{center}
\leavevmode
\psfig{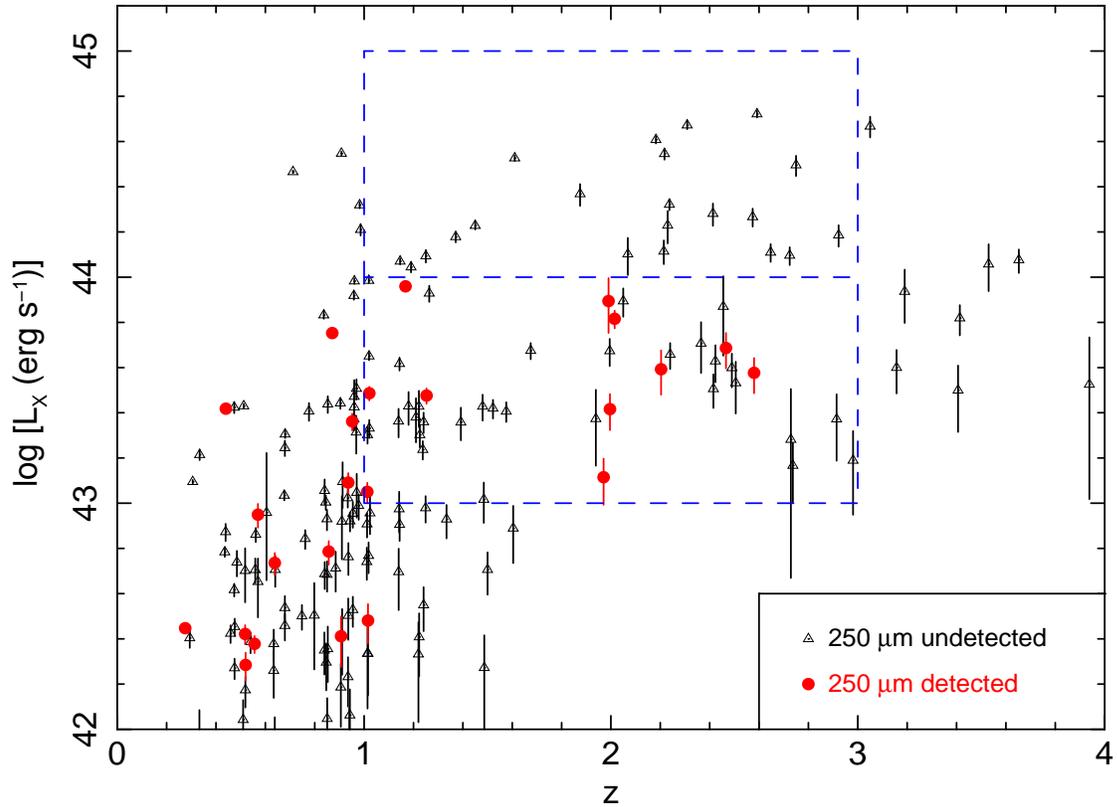}
\caption{Redshifts and 2--8~keV X-ray luminosities of AGN in the CDF-N. The
  luminosities have been K-corrected assuming a spectrum $S_{\nu}\propto
  \nu^{-0.9}$ and are not corrected for intrinsic absorption. The blue dashed
  rectangles delimit the luminosity decades above and below 
  $10^{44}$~erg~s$^{-1}$ 
  in the $1<z<3$ redshift range. Error bars correspond to 68\% confidence.} 
\label{fig:z_Lx}
\end{center}
\end{figure}

The distribution of CDF-N AGN in the
redshift - luminosity plane is shown in Figure \ref{fig:z_Lx}, 
and shows a striking trend of 250~$\mu$m
detectability with X-ray luminosity: of the 24 AGN detected at
250~$\mu$m, none of them have $L_{X} > $
$10^{44}$~erg~s$^{-1}$. 
The redshift range between 1 and
3 is of most interest, because it corresponds to the epoch in which
powerful AGN accreted most of their black hole mass and present-day
massive galaxies formed most of their stars. Within this redshift
range, Fig. \ref{fig:z_Lx} shows that 11 out of 44 AGN
($25^{+8}_{-7}$\%) with
$10^{43}$~erg~s$^{-1} < L_{X} < 10^{44}$~erg~s$^{-1}$ 
are detected
at 250~$\mu$m, while none of the 21 objects with $L_{X} >$
$10^{44}$~erg~s$^{-1}$ are detected. The difference in detection
rates has a significance of $>99\%$ according to a single-tail
Fisher's exact test.  We have considered the effects that
incompleteness in the spectroscopic redshifts, or absorption of the
X-ray flux by gas and dust, might have on our results. We find that
the systematic non-detection of the powerful AGN is robust against
both effects, although X-ray absorption does appear to be a common
property of the 250~$\mu$m-detected AGN. We have also verified the low 250~$\mu$m detection rate of AGN
with $L_{X} >$
$10^{44}$~erg~s$^{-1}$ using the Extended Chandra Deep Field South
field, finding that of 49 such sources with $1<z<3$, only 1 is
detected at 250~$\mu$m.

\begin{table}[h]
\begin{center}
\begin{tabular}{cccccccc}
ID&Redshift&Log $L_{X}$ (erg~s$^{-1}$)& Log $N_{H}$ (cm$^{-2}$) & $N_{H}$ corr & Log $L_{IR}$ (L$_{\odot}$)& AGN (\%) & SFR (M$_{\odot}$ yr$^{-1}$)\\
\hline\\
35  & 2.203 & 43.59$^{+0.08}_{-0.11}$ & $23.6^{+0.1}_{-0.2}$ & $0.15^{+0.05}_{-0.07}$ & $12.70\pm 0.03$ & 12 & 750 -- 850  \\
109 & 2.580 & 43.58$^{+0.07}_{-0.09}$ & $23.4^{+0.1}_{-*}$   & $0.09^{+0.03}_{-0.09}$ & $13.01\pm 0.05$ & 5  & 1660 -- 1750 \\
135 & 2.466 & 43.69$^{+0.07}_{-0.09}$ & $>24.0$              & $>0.30$                & $12.81\pm 0.05$ & 4  & 1060 -- 1110 \\
158 & 1.013 & 43.05$^{+0.04}_{-0.05}$ & $23.1^{+0.1}_{-0.1}$ & $0.15^{+0.02}_{-0.02}$ & $12.29\pm 0.09$ & 4  & 320 -- 330  \\
190 & 2.015 & 43.81$^{+0.04}_{-0.04}$ & $23.6^{+0.1}_{-0.1}$ & $0.16^{+0.02}_{-0.02}$ & $12.88\pm 0.03$ & 21 & 1030 -- 1300 \\
331 & 1.253 & 43.48$^{+0.03}_{-0.04}$ & ---                  & ---                    & $12.51\pm 0.07$ & 5  & 530 -- 550  \\
366 & 1.970 & 43.11$^{+0.08}_{-0.12}$ & $23.4^{+0.1}_{-0.2}$ & $0.11^{+0.03}_{-0.04}$ & $12.84\pm 0.05$ & 3  & 1140 -- 1170 \\
368 & 1.996 & 43.42$^{+0.07}_{-0.09}$ & $>23.8$              & $>0.26$                & $12.41\pm 0.06$ & 3  & 420 -- 430  \\
384 & 1.021 & 43.49$^{+0.03}_{-0.03}$ & $23.4^{+0.1}_{-0.1}$ & $0.25^{+0.02}_{-0.03}$ & $11.55\pm 0.17$ & 11 & 50 -- 60   \\
455 & 1.168 & 43.96$^{+0.02}_{-0.02}$ & ---                  & ---                    & $11.97\pm 0.04$ & 30 & 110 -- 160  \\
500 & 1.990 & 43.89$^{+0.10}_{-0.14}$ & $23.2^{+0.2}_{-*}$   & $0.07^{+0.03}_{-0.07}$ & $12.62\pm 0.03$ & 4  & 690 -- 710  \\
&&&&&&&\\
\end{tabular}
\end{center}
\caption{Properties of SPIRE-detected AGN. Data are given for AGN with $1<z<3$ and $10^{43}$~erg~s$^{-1} < L_{X} < 10^{44}$~erg~s$^{-1}$. The first column gives the ID number of the source in the X-ray catalogue\cite{alexander03}. The second column gives the redshift, and the third column gives the logarithm of the 2--8 keV X-ray luminosity. The fourth column gives the log of the column density of absorbing gas (units of hydrogen atoms per cm$^{2}$) implied by the ratio of 2--8~keV to 0.5--2~keV X-rays; a dash indicates no evidence for photoelectric absorption in X-rays, and an asterix is used where the lower limit to the column density is unconstrained. The fifth column gives the correction to log~$L_{X}$ to account for the absorption. The sixth column gives the 8--1000~$\mu$m IR luminosity. The seventh column gives the maximum likely contribution of an AGN to the infrared luminosity, and the eighth column gives the range of star formation rate implied by the infrared luminosity, where the upper and lower limits correspond to zero AGN contribution and the maximum AGN contribution to the IR luminosity, respectively. Photometry for the spectral energy distributions was extracted from {\em Spitzer} and SPIRE images using the X-ray and 24~$\mu$m catalogue positions as priors\cite{roseboom10}. Total 8--1000~$\mu$m infrared luminosities were then determined by fitting templates\cite{siebenmorgen07} to the spectral energy distributions\cite{symeonidis09}. Upper limits to the AGN contribution to the infrared luminosities were obtained by normalising an AGN template in the mid infrared\cite{seymour11}.
}
\label{tab:luminosities}
\end{table}

Infrared spectral energy distributions for the 250~$\mu$m-detected AGN
were constructed by combining the SPIRE photometry with {\em Spitzer}
3.6 -- 160~$\mu$m photometry. X-ray and infrared properties of the 11
250~$\mu$m-detected AGN with $1<z<3$ and $L_{X}$ in the range
$10^{43}-10^{44}$~erg~s$^{-1}$ are given in Table \ref{tab:luminosities}. In
most cases the AGN contributes less than 10\% to the infrared
luminosity. The best-fit infrared luminosities lie between $4\times 10^{11}$
and $10^{13}$~L$_{\odot}$, implying star formation rates between 50 and
1750 M$_{\odot}$ per year\cite{kennicutt98}.

\begin{figure}
\begin{center}
\leavevmode
\psfig{figure=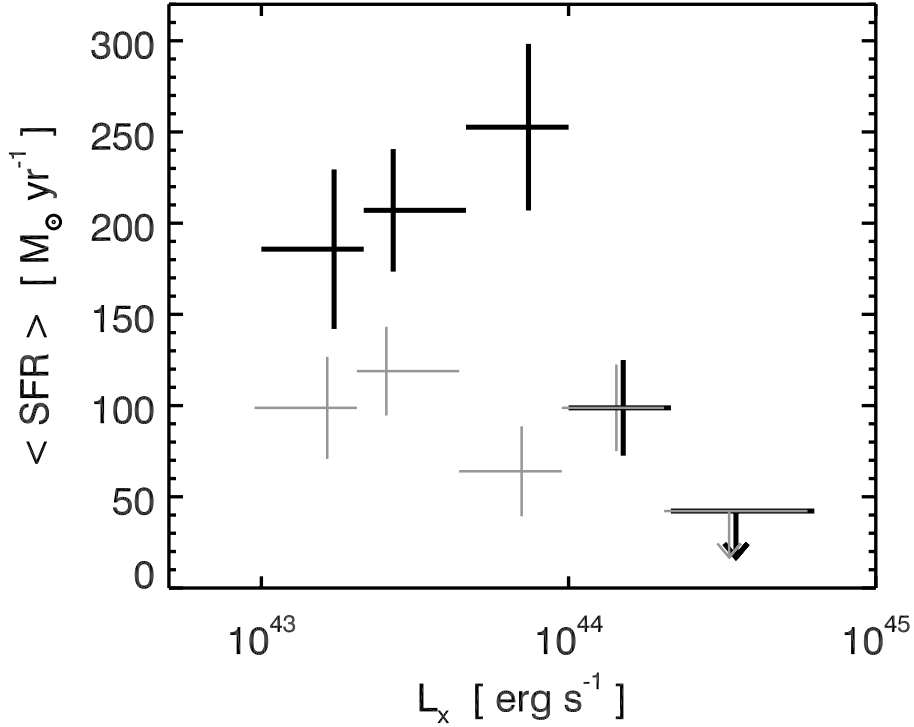,width=120truemm,angle=0}
\caption{Star formation rates derived from averaged far-IR
  luminosities of $1<z<3$ AGN.  We converted the 250, 350 and
  500~$\mu$m flux densities for each source into an equivalent
  8-1000$\mu$m luminosity by fitting a grey-body curve, with a
  temperature of 30~K in the rest-frame of the source, an emissivity
  index of $\beta=1.6$, and a power-law extension to the Wien
  side\cite{blain03} and multiplying by 4$\pi D_{L}^{2}$ where $D_{L}$
  is the luminosity distance. Fluctuations in the map sometimes
  scatter the fluxes of undetected sources to negative values, which
  translate to negative luminosities when multiplied by 4$\pi
  D_{L}^{2}$. Such negative solutions for individual AGN were allowed
  so as not to produce an artificial positive bias in the
  averages. The luminosities were averaged in 5 bins in $L_{X}$, which
  were chosen to include a similar number of AGN in each bin. The
  average luminosities were then converted to star formation
  rates\cite{kennicutt98}.  AGN which are individually detected at
  250~$\mu$m are included in the averages shown in bold black, but have been
  excluded from the averages which are shown in grey, to show the
  contribution that these sources make to the average star formation
  rates. The grey points have been offset horizontally from the bold black 
  points for clarity. Error bars correspond to 68\% confidence and were 
  determined
  by bootstrap resampling, with a 7\% systematic error added in
  quadrature to account for the calibration error on SPIRE photometry.
}
\label{fig:agn_sfr}
\end{center}
\end{figure}

We performed a stacking analysis for the
$1<z<3$ AGN to probe below the confusion limit of the SPIRE images. We
split the sample into five bins of $L_{X}$ from $10^{43}$ to
$10^{45}$~erg~s$^{-1}$ and determined the average star formation rates
of AGN in each bin.  The results are shown in Figure
\ref{fig:agn_sfr}.  In the redshift range $1<z<3$, the mean star
formation rate in AGN with $L_{X}$ of $10^{43}-10^{44}$~erg~s$^{-1}$ is
214$\pm25$~M$_{\odot}$~per~year, compared to a mean star formation
rate for AGN with $L_{X}> 10^{44}$~erg~s$^{-1}$ of
65$\pm18$~M$_{\odot}$~per~year. These averages are independent of
the SPIRE 250~$\mu$m detection limit because they are obtained from a
stack of all sources within a given range of $L_{X}$, whether detected
at 250~$\mu$m or not.

At redshifts of 1--3, the X-ray luminosity of 
10$^{44}$~erg~s$^{-1}$,
which divides the regions of 250~$\mu$m detection and non-detection in
Fig. \ref{fig:z_Lx}, corresponds approximately to the knee in the
luminosity function of AGN\cite{ebrero09}.  The steep shape of the
luminosity function at 
$L_{X} > 10^{44}$~erg~s$^{-1}$ 
implies that
this part of the luminosity function is dominated by objects which are
at the peak of their accretion rates. Our observations therefore imply that
the most prodigious episodes of star formation are common in the host
galaxies of $1<z<3$ AGN, but avoid powerful 
AGN in which accretion is at its peak.

This systematic non-coincidence of the peak periods of star formation
and accretion implies a palpable interaction between the two
processes, and provides a powerful discriminator for the form of AGN
feedback which is responsible for terminating star formation in the
host galaxy. Two families of feedback models have been proposed,
widely referred to as `quasar mode' and `radio mode'\cite{croton06}. 
In quasar-mode feedback, a luminous AGN generates a powerful
wind which terminates star formation by driving the interstellar
medium from the surrounding host galaxy. 
In radio-mode feedback, star
formation is suppressed because collimated jets of relativistic
particles emitted by a radiatively-inefficient AGN prevent gas in 
the surrounding hot halo
from cooling, thereby starving the galaxy of cool gas from which to
form stars.

Radio-mode feedback is commonly invoked in semi-analytical models to
limit galaxy masses and luminosities\cite{bower06,croton06}. In these
models, black holes grow through luminous accretion episodes and black
hole mergers. The correlation between black hole and bulge mass comes
from assuming that a fixed fraction of the gas is accreted by the
nucleus during each star forming episode that results from a galaxy
merger or disc instability, and hence star formation and accretion
rate should be correlated over the full range of luminosity. Our
observations are therefore inconsistent with models
 in which AGN influence their host galaxies
only through radio mode feedback\cite{bower06,croton06}. 
In contrast, models of galaxy
formation in which quasar-mode feedback is responsible for terminating
the star formation\cite{granato04, springel05, dimatteo05, sijacki07}, and 
which have received some observational support 
recently\cite{farrah12, canodiaz12}, 
predict that the AGN luminosity peaks later than the star formation
rate, and thus are consistent with our observations. These models also
predict that residual star formation, at the level of a few tens of per 
cent of the
peak, will continue during the period in which the AGN luminosity is
at its maximum, consistent with our stacked results which show that on
average AGN with 
$L_{X}>10^{44}$~erg~s$^{-1}$ are still forming stars at approximately
65~M$_{\odot}$ per year. 
Our observations do not discriminate between
models invoking major mergers\cite{springel05} or accretion of gas
into a massive halo\cite{granato04} as the trigger for the intense
star formation. After the interstellar medium has been driven out by
the luminous AGN and the AGN itself becomes starved of fuel, radio-mode
feedback is the most credible agent by which further star formation is
inhibited.

\vspace{5mm}
\noindent {\bf Supplementary Information} is linked to the online version of 
the paper at www.nature.com/nature\\

\noindent
{\bf Acknowledgments}
SPIRE has been developed by a consortium of institutes led
by Cardiff Univ. (UK) and including Univ. Lethbridge (Canada);
NAOC (China); CEA, LAM (France); IFSI, Univ. Padua (Italy);
IAC (Spain); Stockholm Observatory (Sweden); Imperial College
London, RAL, UCL-MSSL, UKATC, Univ. Sussex (UK); Caltech,
JPL, NHSC, Univ. Colorado (USA). This development has been
supported by national funding agencies: CSA (Canada); NAOC
(China); CEA, CNES, CNRS (France); ASI (Italy); MCINN (Spain);
SNSB (Sweden); STFC, UKSA (UK); and NASA (USA).\\

\noindent{\bf Author Contributions}
This paper represents the combined work of the HerMES collaboration,
the SPIRE Instrument Team's extragalactic survey. 
M. Page planned the study, and wrote
the draft version of the paper. M. Symeonidis fitted models to the
spectral energy distributions of the sources and J.D. Vieira performed
the stacking analysis. All other coauthors of this paper contributed
extensively and equally by their varied contributions to the SPIRE
instrument, Herschel mission, analysis of SPIRE and HerMES data,
planning of HerMES observations and scientific support of HerMES, and
by commenting on this manuscript as part of an internal review
process.\\

\noindent{\bf Author Information}
Reprints and permissions information is available at
www.nature.com/reprints. The authors declare that they have no
competing financial interests. Correspondence and requests for
materials should be addressed to M.Page (mjp@mssl.ucl.ac.uk).
\newpage

\begin{center}
{\Large \bf 
\noindent
{Supplementary Information}}
\end{center}
\vspace{4mm}

\noindent
{\bf Spectroscopic redshifts of CDF-N X-ray sources}
\vspace{2mm}

We have taken a
recent compilation$^{13}$ 
as our main source of spectroscopic redshifts. These redshifts
are given to two decimal places, but the majority of
them originate from earlier studies, so we have in most cases obtained
more precise redshifts by tracing them back to the original
publications. Where there is a significant discrepancy ($>2$\%)
between a spectroscopic redshift in our main source and that given
in earlier literature, we have adopted the redshift from our 
main source$^{13}$. We have utilised a number of spectroscopic redshifts 
which are not listed there. The redshifts and literature 
references for these cases are given in Table~S1.
\vspace{3mm}

\noindent
{\bf Effect of redshift incompleteness}
\vspace{2mm}

While we have restricted our study to objects within the CDF-N which
have optical spectroscopic identifications and redshifts to ensure
fidelity in redshifts and luminosities, we must examine whether the
limited spectroscopic completeness (64\%) has any effect on our
results. A large fraction of the X-ray sources within the CDF-N have
photometric redshifts (i.e. redshifts estimated by comparing
photometry in a number of bands to that expected from galaxy template
spectra as a function of redshift) in the same compilation that we use 
for our spectroscopic redshifts$^{13}$.
Although these photometric redshifts are of low
accuracy and reliability compared to the spectroscopic redshifts, they
raise the redshift completeness of the 2--8 keV X-ray sources to 87\%,
and so offer a good test as to whether the spectroscopic completeness
has a significant influence on our results. Table~S2
gives statistics equivalent to those shown
in Table 1 of the main paper, but with the photometric redshifts
included. As found for the spectroscopic-only sample, no AGN more luminous 
than $10^{44}$~erg~s$^{-1}$ are associated with 250~$\mu$m sources.

\begin{table}
\singlespacing
\begin{center}
\begin{tabular}{lcl}
Source number in X-ray catalogue$^{12}$&Spectroscopic redshift&Reference\\
\hline\\
91&0.294&\cite{barger02}\\
121&0.520&\cite{barger03}\\
152&0.540&$^{14}$\\ 
191&0.952&$^{14}$\\ 
196&0.680&\cite{cohen96}\\
205&0.943&$^{14}$\\ 
220&4.420&\cite{waddington99}\\
223&2.724&\cite{barger01}\\
243&0.557&\cite{wirth04}\\
266&1.2258&$^{14}$\\ 
287&2.737&$^{14}$\\ 
292&0.497&\cite{cowie04}\\
317&1.760&$^{14}$\\ 
342&0.518&\cite{barger02}\\
377&3.1569&$^{14}$\\ 
412&1.1435&$^{14}$\\ 
423&2.365&$^{14}$\\ 
441&0.634&\cite{barger03}\\
443&0.033&\cite{barger02}\\
470&0.231&\cite{barger03}\\
\hline\\
\end{tabular}
\end{center}
{Table S1: Spectroscopic redshifts which do not come from our main source$^{13}$
and references to the literature in which these redshifts are published.
}
\doublespacing
\end{table}

We can be further reassured that our main result is not affected by
the spectroscopic incompleteness by looking specifically at the
identification statistics of those 2--8 keV X-ray sources which are
associated with 250~$\mu$m sources. A total of 33 2--8 keV X-ray sources
are associated with 250~$\mu$m sources, of which 27 are
spectroscopically identified. The spectroscopically-identified
fraction of these sources is thus 82\%, higher than the 2--8 keV source
population as a whole. For the 6 sources without spectroscopic
redshifts, Fig.~S1
shows the tracks of X-ray
luminosity as a function of redshift.  We can see that with respect to
the $1<z<3$, $10^{43}$\,erg~s$^{-1}$$<L_{X}<10^{45}$\,erg~s$^{-1}$ region which is well sampled
by the X-ray and submillimetre surveys, sources 31, 37 and 431 from
the X-ray catalogue$^{12}$
have $L_{X}<10^{44}$\,erg~s$^{-1}$ for all $1<z<3$. The tracks
for sources 107, 221 and 406 all cross the
upper rectangle corresponding to $1<z<3$, $10^{44}$~erg~s$^{-1}$$<L_{X}<10^{45}$~erg~s$^{-1}$,
and so should be examined in more detail. The track for source 221
only enters this rectangle for a tiny redshift interval
($2.94<z<3.0$). Although our main redshift compilation$^{13}$ does 
not give a photometric
redshift for source 221, a photometric
redshift of 1.9 is found elsewhere in the literature\cite{donley07}, 
in which case it has $L_{X}<10^{44}$~erg~s$^{-1}$ in keeping with 
the other 250~$\mu$m-detected AGN.
Source 107 is within 6 arcsec of the same 250~$\mu$m source as a
brighter X-ray source (source 109), which has a spectroscopic redshift
of 2.58. The brighter X-ray source is closer to the 250~$\mu$m source
position than source 107, and source extraction using X-ray and
24~$\mu$m positions as priors$^{25}$
does not recover 
significant 250~$\mu$m emission for source
107, leaving some doubt about the
reality of the association.  Source 107 has a photometric redshift of
1.15, but could be physically associated with source 109 at $z=2.58$
given that they are only separated by 2.7 arcsec. It can be seen from
Fig.~S1
that whichever of the two redshifts is
adopted for source 107, it has $L_{X}$ between $10^{43}$~erg~s$^{-1}$ and
$10^{44}$~erg~s$^{-1}$, and hence if the association of source 107 and the
250~$\mu$m source is real, it too has $L_{X}<10^{44}$~erg~s$^{-1}$, 
in keeping with 
the other 250~$\mu$m-detected AGN. Finally, we come
to source 406, which is the brightest of the six X-ray sources. From
the track in Fig.~S1,
it appears the most likely of the six to
be found within the $1<z<3$, $10^{44}$~erg~s$^{-1}$$<L_{X}<10^{45}$~erg~s$^{-1}$ 
region of our
survey. This source is extremely faint in the optical, with an $R$-band
magnitude of 27.8 \cite{barger03}, and there are no estimates of
photometric redshift available in the literature. The association of
this X-ray source with the 250~$\mu$m source must be regarded with some doubt, 
however, because it is located just 5.5 arcsec to the south of a
brighter 24~$\mu$m and radio source which is closer to the 250~$\mu$m
position. Source extraction using X-ray and 24~$\mu$m positions as
priors$^{25}$
detects source 406 only at the 3$\sigma$
level. Nonetheless, this source does represent the single example of
an X-ray source, potentially associated with a 250~$\mu$m source, with
a reasonable likelihood of lying within the $1<z<3$, 
$L_{X} >10^{44}$~erg~s$^{-1}$
region. Equally, it may fall within the $1<z<3$, $L_{X} <10^{44}$~erg~s$^{-1}$
region.
To summarise, it appears that the sources without spectroscopic
redshifts are likely to harbour at most one object within the 
redshift range $1<z<3$, which has $L_{X}>10^{44}$~erg~s$^{-1}$.

\begin{table}
\begin{center}
\begin{tabular}{lcccc}
Region of (z,L) parameter space&Number&Number of AGN&Expected number of&Fraction of AGN \\
&of AGN&associated with&spurious associations&associated with\\
&&250~$\mu$m sources&&250~$\mu$m sources\\
\hline\\
all $z$,\ \ \ \ \ \ \ \ 
$10^{42} < L_{X} < 10^{45}$~erg~s$^{-1}$&248&27&2.8&11$\pm$2 ($\pm$4) \%\\
$1<z<3$, $10^{43} < L_{X} < 10^{44}$~erg~s$^{-1}$&61&12&0.7&20$\pm$6 ($_{-9}^{+12}$) \%\\
$1<z<3$, $10^{44} < L_{X} < 10^{45}$~erg~s$^{-1}$&25&0&0.3&$<4$ ($<11$) \%\\
\hline\\

\end{tabular}
\end{center}
{Table S2: 250~$\mu$m detection statistics in the various parts of parameter
  space when AGN with photometric redshifts are included as well as those 
with spectroscopic redshifts. The first row corresponds to the full AGN 
luminosity and redshift parameter space in the CDF-N, 
  while the second and third rows correspond to the
  regions enclosed within the blue dashed lines in Fig. 1.
}
\end{table}

\begin{figure}
\begin{center}
\leavevmode
\psfig{figure=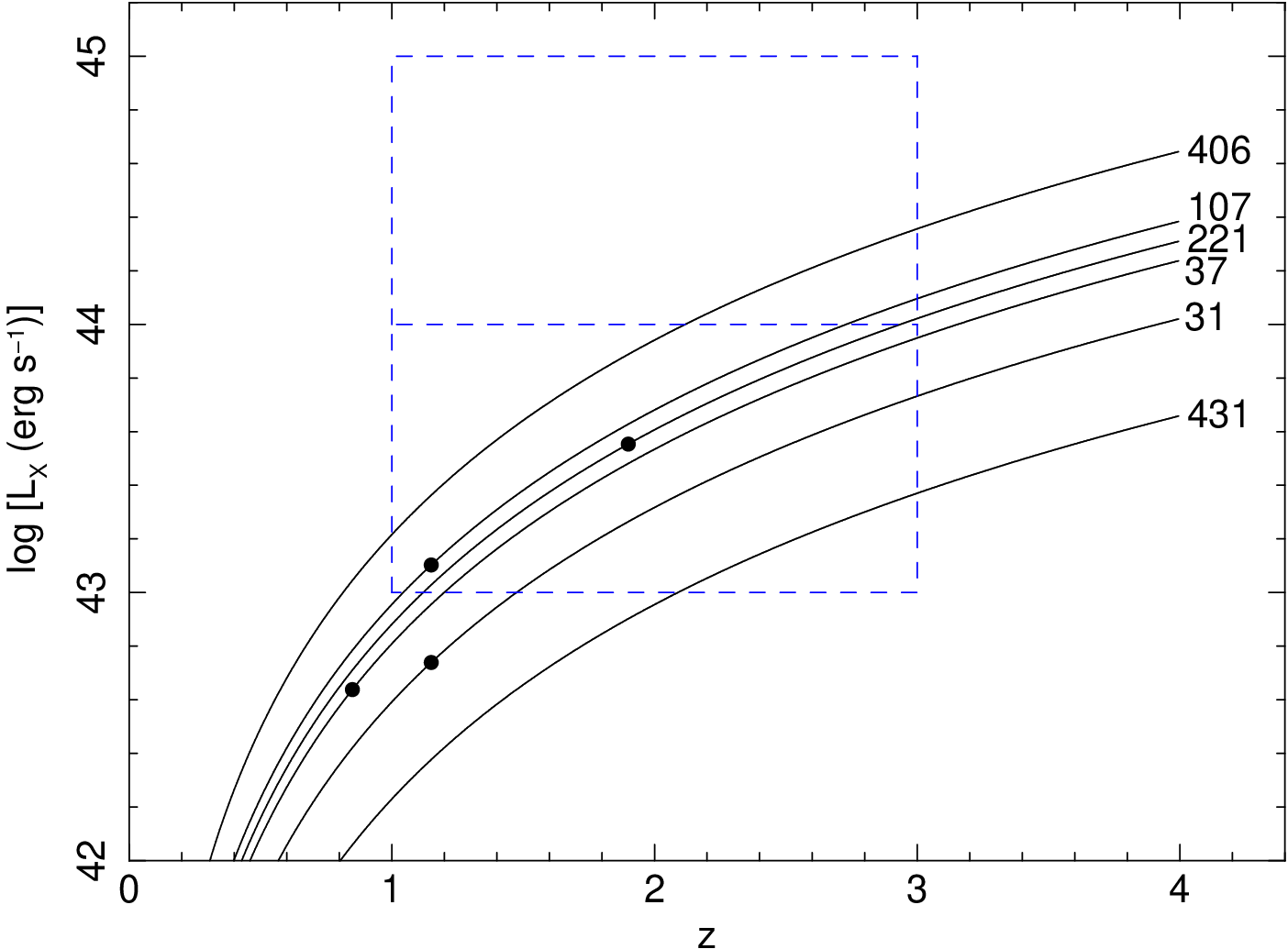,width=130truemm,angle=270}
\end{center}
{Figure S1: 2--8 keV X-ray luminosities as a function of redshift for 2--8
  keV X-ray sources which are associated with 250~$\mu$m sources and
  which do not have spectroscopic redshifts. The bold dots indicate 
photometric redshfits, where available. The blue dashed rectangle
  delimits the luminosity decades above and below 
  $10^{44}$\,erg~s$^{-1}$ 
  in the
  $1<z<3$ redshift range as in Fig.~1.
Error bars correspond 
  to 68\% confidence.}
\end{figure}

\vspace{3mm}

\noindent
{\bf Effect of X-ray absorption}
\vspace{2mm}

X-rays in the 2--8 keV band can penetrate considerable column densities
of material, but column densities $>10^{23}$cm$^{-2}$ can lead to
significant attenuation of the X-ray flux. Conceivably, the different
250~$\mu$m detection rates for AGN above and below $10^{44}$\,erg~s$^{-1}$
could be a consequence of X-ray absorption, if 250~$\mu$m-luminous AGN
are systematically more heavily absorbed than AGN which are not
detected at 250~$\mu$m. X-rays in the 0.5--2 keV band are
absorbed more easily than those in the 2--8 keV band which was used to
select our X-ray sample, and hence the hardness ratio of 2--8
keV / 0.5--2 keV X-rays is a crude but effective estimator of the degree
of X-ray absorption. 

\begin{figure}
\begin{center}
\leavevmode
\psfig{figure=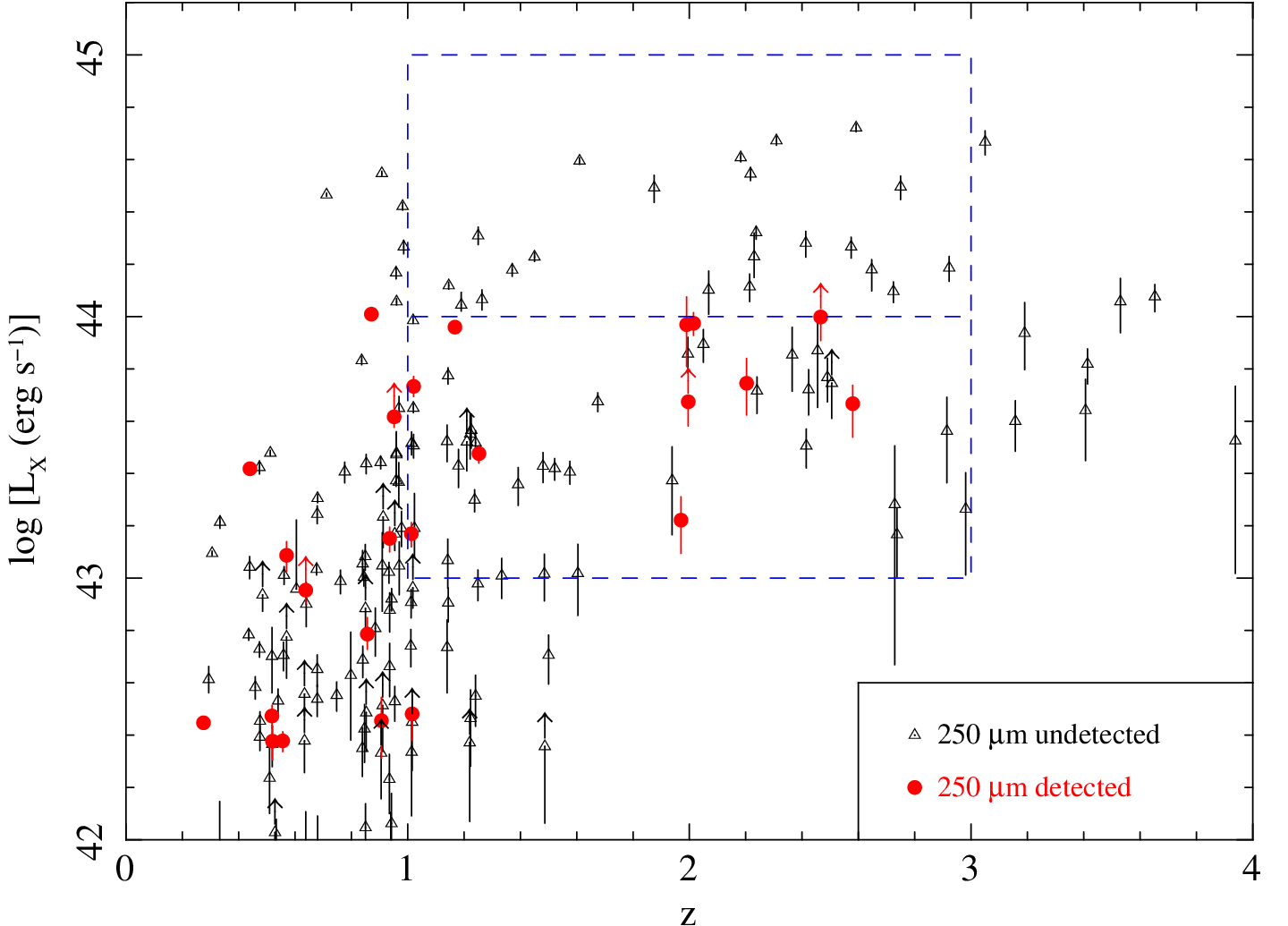,width=130truemm,angle=270}
\end{center}
{Figure S2: 2--8 keV X-ray luminosities as a function of redshift after 
correction for absorption. AGN with lower-limit hardness
ratios have absorption-corrected luminosities with unconstrained upper
bounds, and these are shown as upward pointing arrows. 
The blue dashed rectangle
  delimits the luminosity decades above and below 
  $10^{44}$\,erg~s$^{-1}$ 
  in the
  $1<z<3$ redshift range as in Fig.~1.
Error bars correspond 
  to 68\% confidence.}
\end{figure}

For each 2--8 keV X-ray source, we used the Portable Interactive
Multi-Mission Simulator
(PIMMS; {\newline http://heasarc.nasa.gov/docs/software/tools/pimms.html)}
to convert from hardness ratios to spectral indices and column
densites.  The intrinsic spectra of faint AGN without absorption are
power laws of the form $S_{\nu}\propto \nu^{-\alpha}$ with a mean
$\alpha=0.9$, and an rms scatter in $\alpha$ of 0.2$^{15}$.
We took a hardness ratio corresponding to
$\alpha=0.5$ ($2\sigma$ from the mean) as the threshold beyond which
we considered sources as absorbed. For absorbed sources, we used PIMMS
to estimate the column density, assuming an intrinsic spectral index
$\alpha=0.9$. PIMMS was then used to calculate the expected degree of
attenuation to the observed 2--8 keV X-rays from the column density
estimate. The X-ray luminosities were then adjusted accordingly. Where
only a lower limit to the hardness ratio is available (due to the
faintness of the object in the 0.5--2 keV X-ray band), the luminosity
correction corresponding to the lower limit is
applied. 

Fig.~S2
shows how Fig.~1
would be
affected by the absorption corrections.  More than half of the
250~$\mu$m-detected AGN increase in luminosity when the absorption
corrections are applied, but on the whole these corrections are quite
modest, with the majority amounting to less than 40\% change in
luminosity. None of the 250~$\mu$m-detected AGN move into the $1<z<3$,
$10^{44}$\,erg~s$^{-1}$~$<L_{X}<10^{45}$\,erg~s$^{-1}$ region.
Although several of the 250~$\mu$m-detected AGN move close to the
$10^{44}$\,erg~s$^{-1}$ boundary, and the absorption corrections for
two of the 250~$\mu$m-detected AGN are lower limits such that they may
have intrinsic luminosities $>10^{44}$\,erg~s$^{-1}$, the essential
characteristic of Fig.~1, the low 250~$\mu$m detection rate above
$10^{44}$\,erg~s$^{-1}$, persists.  \vspace{3mm}

The fraction of absorbed AGN is known to depend on luminosity\cite{dellaceca08} 
such that absorption is less common in the most
luminous AGN. In light of this interdependency between luminosity and
absorption, we have also examined the 250~$\mu$m detection rate of AGN
with $1<z<3$ and $10^{43}$~erg~s$^{-1}$~$<L_{X}<10^{45}$~erg~s$^{-1}$
as a function of X-ray absorption rather than X-ray luminosity.
Dividing the sample into absorbed and unabsorbed sources using the
hardness-ratio criterion given above, we have 9 absorbed and 2
unabsorbed AGN detected at 250~$\mu$m, and 27 each of absorbed and
unabsorbed sources which are not detected at 250~$\mu$m. Applying a
Fisher's exact test, we find that the difference in 250~$\mu$m
detection rate between absorbed and unabsorbed sources is significant
at a little less than 95\%. We can reduce the influence of the
intrinsic anti-correlation between absorption and luminosity within
the sample under test by excluding the sources with X-ray luminosities
above $10^{44}$~erg~s$^{-1}$. We then have 9 absorbed and 2 unabsorbed
sources detected at 250~$\mu$m, with 19 absorbed and 14 unabsorbed
sources undetected at 250~$\mu$m.  Applying the Fisher's test, the
statistical significance at which the 250~$\mu$m detection rate
depends on X-ray absorption drops to 90\%. Hence, whereas we can
demonstrate that the 250~$\mu$m detection rate depends on X-ray
luminosity, we can not state with such confidence that the 250~$\mu$m
detection rate depends on X-ray absorption.

\begin{figure}
\begin{center}
\leavevmode
\psfig{figure=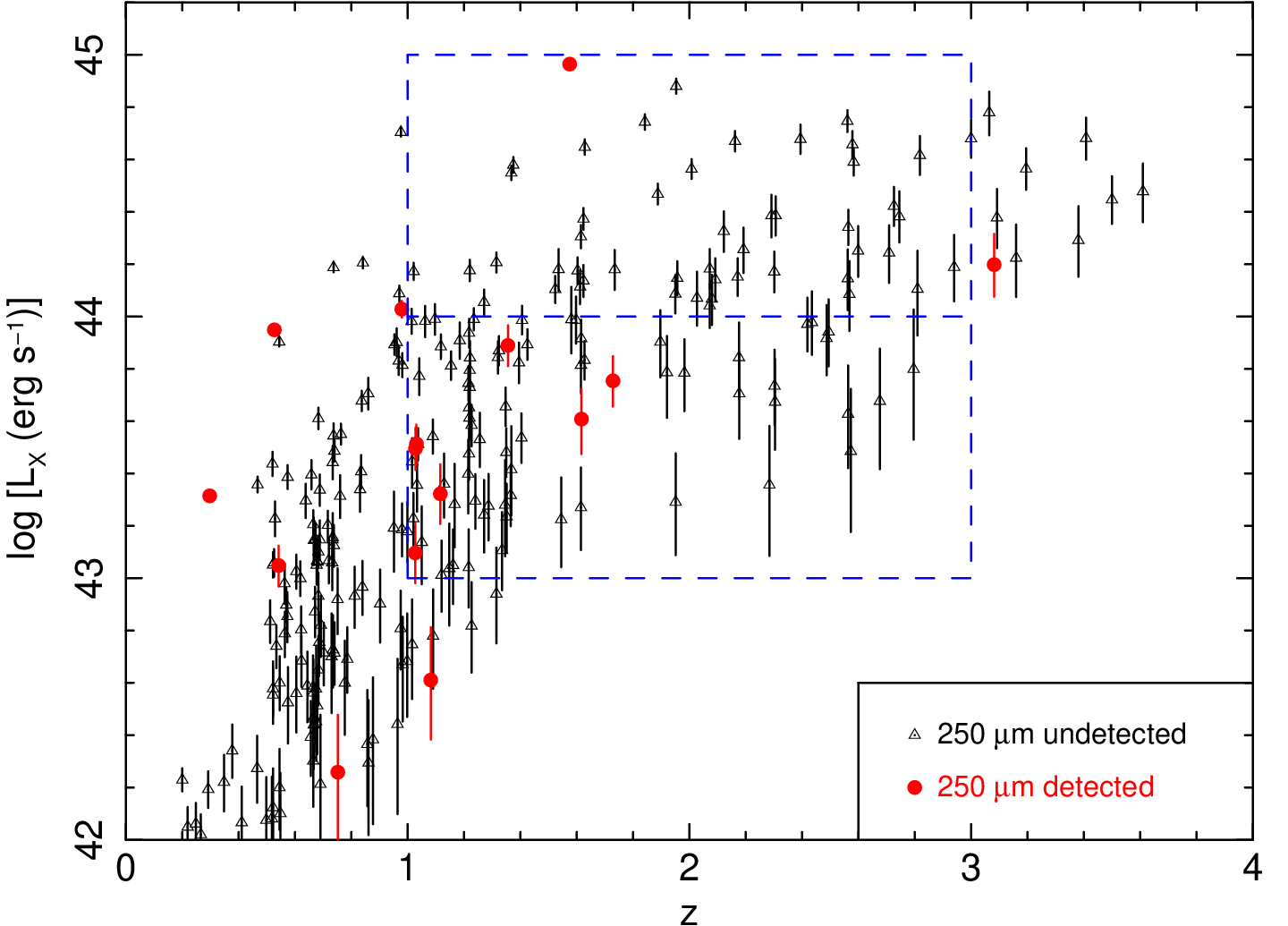,width=130truemm,angle=270}
\end{center}
{Figure S3: 2--8 keV X-ray luminosities as a function of redshift 
for the ECDF-S field. The blue dashed rectangle
  delimits the luminosity decades above and below 
  $10^{44}$~erg~s$^{-1}$ 
  in the
  $1<z<3$ redshift range as in Fig.~1.
Error bars correspond 
  to 68\% confidence.}
\end{figure}

\vspace{3mm}
\noindent
{\bf Verification in the Extended Chandra Deep Field South}
\vspace{2mm}

We have verified that the main result of this work, the paucity of
250~$\mu$m-detections amongst X-ray-luminous AGN, is reproduced in the
Extended Chandra Deep Field South (ECDF-S), which is a larger field
than the CDF-N, containing a larger sample of AGN with
$L_{X}>10^{44}$\,erg~s$^{-1}$. We used the published X-ray 
catalogue\cite{lehmer05} and spectroscopic redshifts\cite{silverman10}.  The
SPIRE observations of the ECDF-S are shallower than those of CDF-N,
but nonetheless are limited primarily by confusion.  As with CDF-N, we
have adopted a detection threshold of 18~mJy at 250~$\mu$m, and
cross-correlated the X-ray and submillimetre catalogues with a
matching radius of 6 arcseconds.  The 250~$\mu$m detections in the
ECDF-S $(z,L)$ plane are shown in Figure~S3,
analogous to Fig. 1 of the main article. Of 49 sources with $1<z<3$
and $L_{X}>10^{44}$~erg~s$^{-1}$, only 1 is detected at 250~$\mu$m,
confirming the low detection rate obtained in the CDF-N.
\vspace{3mm}

\begin{figure}
\begin{center}
\leavevmode
\psfig{figure=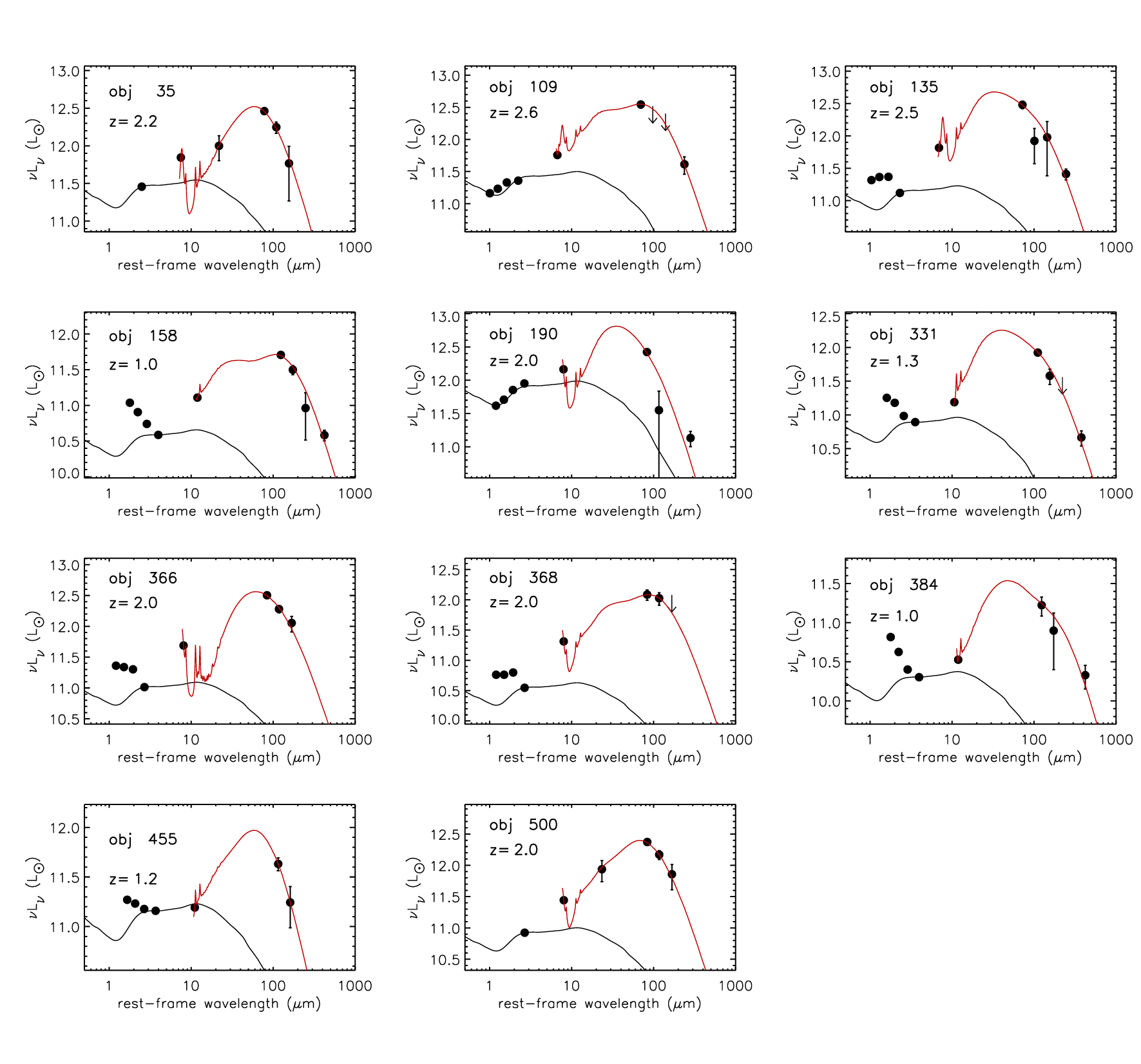,width=140truemm}
\end{center}
{Figure S4: Infrared spectral energy distributions of 250~$\mu$m-detected
  AGN with $1<z<3$ and $10^{43}$~erg~s$^{-1} < L_{X} <
  10^{44}$~erg~s$^{-1}$, based on SPIRE and {\em Spitzer} data. We
included 850~$\mu$m and 1.3~mm photometry from ground-based telescopes
where available\cite{chapman05, tacconi06}.  The black datapoints 
are the measured
  photometry, and the red lines show the star-forming model
  templates$^{26}$
which best fit the data at 24~$\mu$m
  and longward. The black lines show AGN templates\cite{elvis94},
  normalised in the mid-IR to estimate the AGN contribution$^{28}$.
Each panel is labelled
  with the identity of the object in the X-ray catalogue$^{12}$.
}
\end{figure}

\noindent
{\bf Infrared spectral energy distributions of 250~$\mu$m-detected AGN}
\vspace{2mm}

The spectral energy distributions and best fit models for the AGN
associated with 250~$\mu$m sources with $1<z<3$ and
$10^{43}$~erg~s$^{-1} < L_{X} < 10^{44}$~erg~s$^{-1}$, are shown in
Fig.~S4
\vspace{3mm}

\noindent
{\bf Obtaining the average star formation rates through stacking}
\vspace{2mm}

The average star formation rates shown in Fig.~2 were obtained by
averaging the far-infrared luminosities of the AGN to account for the
different redshifts of the different AGN. It is more common in
stacking analyses to average the fluxes of the objects, and then to
use the average flux and the median redshift of the sample to infer an average
luminosity. Here we examine how the results would change if we used this latter method rather than stacking directly in luminosity. Fig.~S5 shows the star formation rates obtained by the two different methods. The two methods give consistent results, and in both cases the average star formation rates are significantly higher for AGN with $\log L_{X} < 10^{44}$~erg~s$^{-1}$ than for AGN with $\log L_{X} > 10^{44}$~erg~s$^{-1}$. We have also examined how the star formation rates would differ if only one of the SPIRE bands were used, instead of all three, to determine the luminosities. For each of the SPIRE bands, we derived a luminosity for each source by normalising a 30~K grey-body in the rest frame to the SPIRE flux and multiplying by 4$\pi D_{L}^{2}$, before averaging the luminosities in each bin. The star formation rates so derived from the three SPIRE bands are shown in Fig.~S5, and are consistent with each other as well as those shown in Fig.~2.

\begin{figure}
\begin{center}
\leavevmode
\psfig{figure=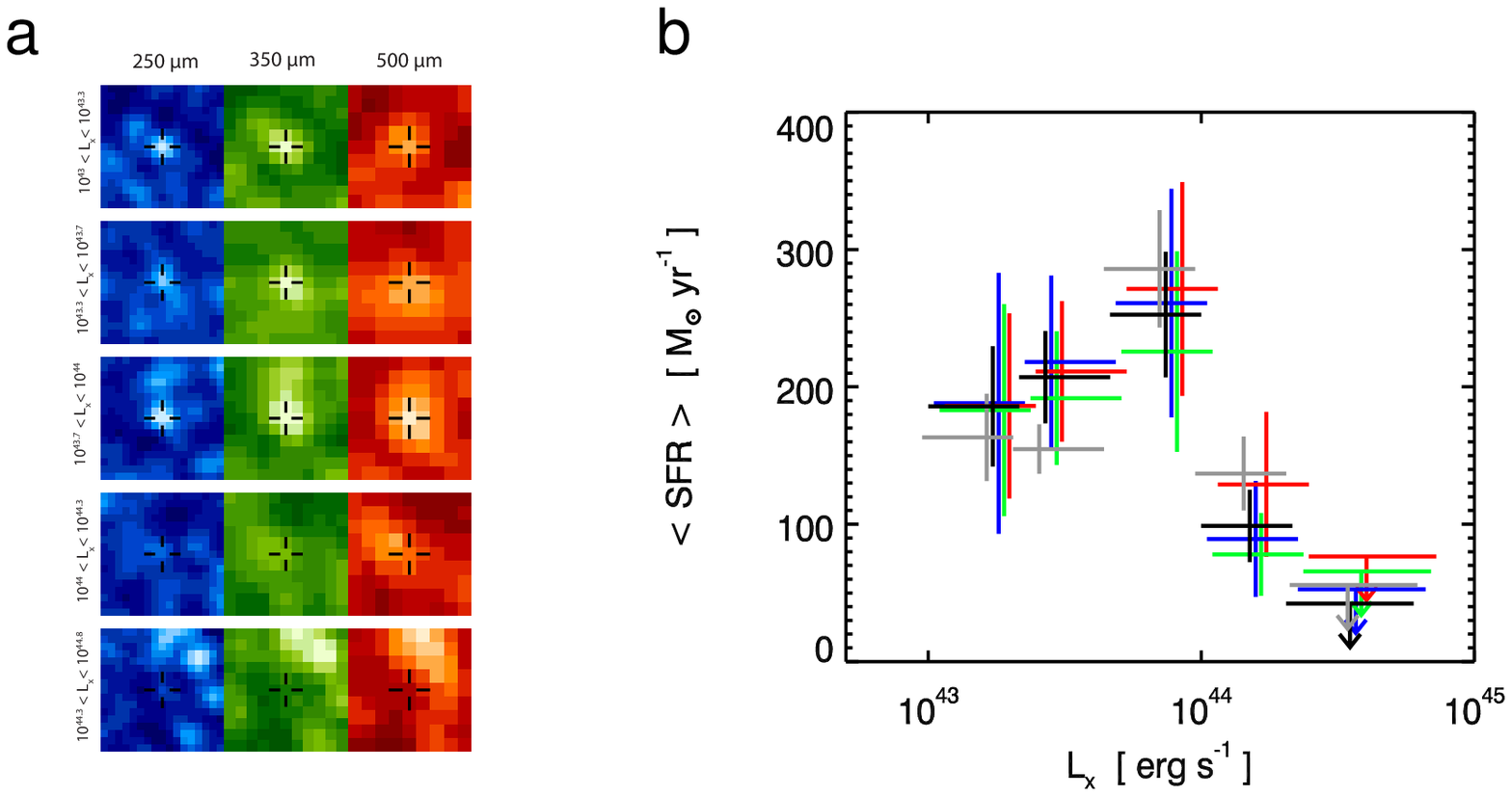,width=140truemm}
\end{center}
{Figure S5: (a) Stacked 250, 350 and 500 $\mu$m SPIRE images of AGN in
  the five luminosity bins used for Fig.~2. The images are arranged in
  increasing wavelength from left to right and decreasing X-ray
  luminosity from bottom to top. (b) Star formation rates obtained
  from the stacked fluxes and median redshifts (grey) compared to the
  star formation rates obtained from stacking the luminosities as in
  Fig.~2 (black). The blue, green and red points show the star
  formation rates obtained in a similar fashion to the black points,
  but in these cases using only one of the SPIRE bands to determine
  the luminosities prior to averaging. Blue, green and red correspond
  to 250, 350 and 500$\mu$m bands. The points have been offset
  slightly in the horizontal direction to aid clarity.  }
\end{figure}


\begin{thebibliography}{}

\bibitem[1]{haring04}
H\"aring \& Rix,
On the Black Hole Mass-Bulge Mass Relation.
{\it Astrophys. J.}, 604, L89-L92 (2004)

\bibitem[2]{silk98}
Silk J. \& Rees M.J., 
Quasars and galaxy formation.
{\it Astron. Astrophys.}, 331, L1-L4 (1998)

\bibitem[3]{fabian99}
Fabian, A.C., 
The obscured growth of massive black holes.
{\it Mon. Not. R. Astron. Soc.}, 308, L39-L43 (1999)

\bibitem[4]{king10}
King, A.R., 
Black hole outflows.
{\it Mon. Not. R. Astron. Soc.}, 402, 1516-1522 (2010)

\bibitem[5]{brandt05}
Brandt W.N. \& Hasinger G.,
Deep Extragalactic X-Ray Surveys.
{\it Annu. Rev. Astron. Astrophys.}, 43, 827-859 (2005)

\bibitem[6]{sanders96}
Sanders D.B. \& Mirabel I.F.,
Luminous Infrared Galaxies.
{\it Annu. Rev. Astron. Astrophys.}, 34, 749-792 (1996)

\bibitem[7]{dimatteo05}
Di Matteo T., Springel V. \& Hernquist L., 
Energy input from quasars regulates the growth and activity of black 
holes and their host galaxies.
{\it Nature}, 433, 604-607 (2005)

\bibitem[8]{springel05}
Springel V., Di Matteo T., \& Hernquist L.,
Modelling feedback from stars and black holes in galaxy mergers.
{\it Mon. Not. R. Astron. Soc.}, 361, 776-794 (2005)

\bibitem[9]{sijacki07}
Sijacki D., Springel V., Di Matteo T., Herhquist L.,
A unified model for AGN feedback in cosmological simulations of 
structure formation.
{\it Mon. Not. R. Astron. Soc.}, 380, 877-900 (2007)

\bibitem[10]{hatziminaoglou10}
Hatziminaoglou, E., et~al.,
HerMES: Far infrared properties of known AGN in the HerMES fields.
{\it Astron. Astrophys.}, 518, L33 (2010)

\bibitem[11]{alexander03} 
Alexander D.M., Bauer F.E., Brandt W.N., et~al., 
The Chandra Deep Field North Survey. XIII. 2 Ms Point-Source Catalogs.
{\it Astron. J.}, 126, 539-574 (2003)

\bibitem[12]{trouille08}
Trouille L., Barger A.J., Cowie L.L., Yang Y., Mushotzky R.F., 
The OPTX Project. I. The Flux and Redshift Catalogs for 
the CLANS, CLASXS, and CDF-N Fields.
{\it Astrophys. J. Suppl.}, 179: 1-18 (2008)

\bibitem[13]{barger08}
Barger A.J., et~al.,
A Highly Complete Spectroscopic Survey of the GOODS-N Field.
{\it Astrophys. J.}, 689, 687-708 (2008)

\bibitem[14]{mateos05}
Mateos S., Barcons X., Carrera F.J., Ceballos M.T., Hasinger G., 
Lehmann I., Fabian A.C., Streblyanska A., 
XMM-Newton observations of the Lockman Hole IV: spectra of the brightest AGN.
{\it Astron. Astrophys.}, 444, 79-99 (2005)

\bibitem[15]{griffin10}
Griffin M.J., et~al.,
The Herschel-SPIRE instrument and its in-flight performance.
{\it Astron. Astrophys.}, 518, L3 (2010)

\bibitem[16]{oliver10}
Oliver, S.J., et~al.,
HerMES: SPIRE galaxy number counts at 250, 350, and 500 $\mu$m.
{\it Astron. Astrophys.}, 518, L21 (2010)

\bibitem[17]{smith10}
Smith A.J., et~al., 
HerMES: point source catalogues from deep Herschel-SPIRE observations.
{\it Mon. Not. R. Astron. Soc.}, 419, 377-389 (2012)

\bibitem[18]{kennicutt98}
Kennicutt, R.C., 
The global Schmidt law in star-forming Galaxies.
{\it Astrophys. J.}, 498, 541-552 (1998)

\bibitem[19]{ebrero09}
Ebrero J., et~al.,
The XMM-Newton serendipitous survey. VI. The X-ray luminosity function.
{\it Astron. Astrophys.}, 493, 55-69 (2009)

\bibitem[20]{croton06}
Croton D.J., et~al., 
The many lives of active galactic nuclei: cooling flows, black holes and the 
luminosities and colours of galaxies.
{\it Mon. Not. R. Astron. Soc.}, 365, 11-28 (2006)

\bibitem[21]{bower06}
Bower R. G., et~al., 
Breaking the hierarchy of galaxy formation.
{\it Mon. Not. R. Astron. Soc.}, 370, 645-655 (2006)

\bibitem[22]{granato04}
Granato G.L., et~al., 
A physical model for the coevolution of QSOs and their spheroidal hosts.
{\it Astrophys. J.}, 600, 580-594 (2004)

\bibitem[23]{farrah12}
Farrah D., et~al.,
Direct evidence for termination of obscured star formation by radiatively 
driven outflows in reddened QSOs.
{\it Astrophys. J.}, 745, 178 (2012)

\bibitem[24]{canodiaz12}
Cano-D\'iaz M., Maiolino R., Marconi A., Netzer H., Shemmer O., Cresci G., 
Observational evidence of quasar feedback quenching star formation at high redshift.
{\it Astron. Astrophys.}, 537, L8 (2012)

\bibitem[25]{roseboom10}
Roseboom I.G., et~al., 
The Herschel Multi-Tiered Extragalactic Survey: source extraction and 
cross-identifications in confusion-dominated SPIRE images.
{\it Mon. Not. R. Astron. Soc.}, 409, 48-65 (2010)

\bibitem[26]{siebenmorgen07}
Siebenmorgen R. \& Kru\"gel E., 
Dust in starburst nuclei and ULIRGs. SED models for observers.
{\it Astron. Astrophys.}, 461, 445-453 (2007)

\bibitem[27]{symeonidis09}
Symeonidis M., et al., 
The link between SCUBA and Spitzer: cold galaxies at z $\le$ 1.
{\it Mon. Not. R. Astron. Soc.}, 397, 1728-1738 (2009)

\bibitem[28]{seymour11}
Seymour, N., et~al., 
HerMES: SPIRE emission from radio-selected active galactic nuclei.
{\it Mon. Not. R. Astron. Soc.}, 413, 1777-1786 (2011)

\bibitem[29]{blain03}
Blain, A.W., Barnard V.E. \& Chapman S.C.,
Submillimetre and far-infrared spectral energy distributions of galaxies: the 
luminosity-temperature relation and consequences for photometric redshifts.
{\it Mon. Not. R. Astron. Soc.}, 338, 733-744 (2003)

\end{thebibliography}

\begin{thebibliography}{}

\bibitem[30]{barger02}
Barger A.J., et~al.,
X-Ray, Optical, and Infrared Imaging and Spectral Properties of the 1 
Ms Chandra Deep Field North Sources.
{\it Astron. J.}, 124, 1839-1885 (2002)

\bibitem[31]{barger03}
Barger A.J., et~al.,
Optical and Infrared Properties of the 2 Ms Chandra Deep Field 
North X-Ray Sources.
{\it Astron. J.}, 126, 632-665 (2003)

\bibitem[32]{cohen96}
Cohen J.G., et~al.,
Redshift Clustering in the Hubble Deep Field.
{\it Astrophys. J.}, 471, L5-L9 (1996)

\bibitem[33]{waddington99}
Waddington I., et~al., 
NICMOS Imaging of the Dusty Microjansky Radio Source 
VLA~J123642+621331 at z=4.424.
{\it Astrophys. J.}, 525, L77-L80 (1999)

\bibitem[34]{barger01}
Barger A.J., et~al.,
Supermassive Black Hole Accretion History Inferred from a Large Sample of 
Chandra Hard X-Ray Sources.
{\it Astron. J.}, 122, 2177-2189 (2001)

\bibitem[35]{wirth04}
Wirth G.D., et~al., 
The Team Keck Treasury Redshift Survey of the GOODS-North Field.
{\it Astron. J.}, 127, 3121-3136 (2004)
 
\bibitem[36]{cowie04}
Cowie L.L., et~al., 
A Large Sample of Spectroscopic Redshifts in the ACS-GOODS Region of the 
Hubble Deep Field North.
{\it Astron. J.}, 127, 3137-3145 (2004)

\bibitem[37]{donley07}
Donley J.L., Rieke G.H., P\/erez-Gonz\/alez P.G., Rigby J.R., 
Alonso-Herrero A., 
Spitzer Power-Law Active Galactic Nucleus Candidates in the 
Chandra Deep Field-North.
{\it Astrophys. J.}, 660, 167-190 (2007)

\bibitem[38]{dellaceca08}
Della Ceca R., et~al., 
The cosmological properties of AGN in the XMM-Newton Hard Bright Survey.
{\it Astron. Astrophys.}, 487, 119-130 (2008)

\bibitem[39]{lehmer05}
Lehmer B.D., et~al., 
The Extended Chandra Deep Field-South Survey: Chandra Point-Source Catalogs.
{\it Astrophys. J. Suppl.}, 161, 21-40 (2005)

\bibitem[40]{silverman10}
Silverman J.D., et~al.,
The Extended Chandra Deep Field-South Survey: Optical Spectroscopy of 
Faint X-ray Sources with the VLT and Keck.
{\it Astrophys. J. Suppl.}, 191, 124-142 (2010)

\bibitem[41]{chapman05}
Chapman S., Blain A.W., Smail I., Ivison R.J.,
A redshift survey of the submillimetre galaxy population.
{\it Astrophys. J.}, 622, 772-796 (2005)

\bibitem[42]{tacconi06}
Tacconi L.J., et~al.,
High-resolution millimetre imaging of submillimeter galaxies.
{\it Astrophys. J.}, 640, 228-240

\bibitem[43]{elvis94}
Elvis M., et al., 
Atlas of quasar energy distributions.
{\it Astrophys. J. Suppl.}, 95, 1-68 (1994)

\end{thebibliography}
\end{document}